This is a pre-print version of the chapter published in *Ethics in Artificial Intelligence: Bias, Fairness and Beyond*. The final authenticated version is available online at https://link.springer.com/chapter/10.1007/978-981-99-7184-8_4.

To cite this chapter:

Makhortykh M. (2023). No AI After Auschwitz? Bridging AI and Memory Ethics in the Context of Information Retrieval of Genocide-Related Information. In A. Mukherjee, J. Kulshrestha, A. Chakraborty, and S. Kumar (Eds.), *Ethics in Artificial Intelligence: Bias, Fairness and Beyond* (pp. 71-85). Springer. DOI: https://doi.org/10.1007/978-981-99-7184-8_4.

**No AI After Auschwitz? Bridging AI and Memory Ethics in the Context of Information Retrieval of Genocide-Related Information**

Mykola Makhortykh[1]

**Abstract**: The growing application of artificial intelligence (AI) in the field of information retrieval (IR) affects different domains, including cultural heritage. By facilitating organisation and retrieval of large volumes of heritage-related content, AI-driven IR systems inform users about a broad range of historical phenomena, including genocides (e.g. the Holocaust). However, it is currently unclear to what degree IR systems are capable of dealing with multiple ethical challenges associated with the curation of genocide-related information. To address this question, this chapter provides an overview of ethical challenges associated with the human curation of genocide-related information using a three-part framework inspired by Belmont criteria (i.e. curation challenges associated with respect for individuals, beneficence and justice/fairness). Then, the chapter discusses to what degree the above-mentioned challenges are applicable to the ways in which AI-driven IR systems deal with genocide-related information and what can be the potential ways of bridging AI and memory ethics in this context.

## 1 Introduction

Information retrieval (IR) is one of the computer science fields that is closely connected to the developments in the domain of artificial intelligence (AI). Defined as the process of selecting items that are deemed relevant for the user information needs based on the user input [1], IR has been argued to be a particularly promising area of applying AI [2, 3]. The integration of AI can benefit different aspects of IR, ranging from knowledge representation to content indexing and matching [6] to relevance modelling [3, 4]. Consequently, there is a long history of research on AI-driven IR applications, starting with rule-based approaches in the 1980s [2] and ending with the neutral network-based approaches discussed in the 2020s [4].

The importance of AI-driven IR systems has been increasing due to the growth in the amount of information available online. Often referred to as an information overload [5], this phenomenon prompted the need for advanced IR mechanisms for satisfying individual information needs, which are capable of not only processing the large volumes of available information but also recognising the diverse spectrum of user needs and in some cases predicting these needs. Such mechanisms demonstrated their usefulness in multiple domains ranging from healthcare [7, 8] to journalism [9, 10] to e-commerce [11, 12].

This chapter focuses on one particular domain in which AI-driven IR systems are increasingly employed, which is cultural heritage. By facilitating the organisation of heritage-related content both within heritage institutions (e.g. archives [13] or museums [14]), and commercial platforms (e.g. web search engines [15] or social media news feeds [16]), AI-driven IR systems help their users become informed about a broad range of historical phenomena, including genocides such as the Holocaust or Rwanda genocide. Under the condition of a high degree of autonomy, these systems become non-human curators of genocide-related information, which shape how individuals and societies are informed about the past and present atrocities.

Despite the importance of IR systems for curating information about historical and recent genocides, there are multiple concerns about their potential impact on genocide remembrance. The usual concerns about the lack of transparency of AI-driven IR systems are further amplified by the possibility of such non-transparency facilitating manipulations of IR systems, which can potentially interfere with the moral obligations of safeguarding the dignity of genocide victims [17]. Furthermore, such manipulations can facilitate the instrumentalisation of memories about past violence, which can be used for justifying the present stigmatisation, as in the case of the Rohingya persecution in Myanmar [18] or the Russian-Ukrainian war [19].

Besides the above-mentioned concerns about the use of IR for curating genocide-related information, there are also other ethical challenges which have for long been discussed in the context of human curation of historical information, such as the importance of protecting the privacy of individuals [20] or preventing unfair practices of information curation [21]. However, despite the growing body of work concerning the ethics of human curation of genocide-related information [22–24], the capabilities of IR systems to deal with complex ethical issues arising in the context of genocide-related information as well as the perspectives of bridging memory ethics and IR design currently remain under-investigated.

To address this gap, the chapter aims to examine whether the concerns about the human curation of genocide-related information are applicable to AI-driven IR systems and how these concerns can potentially be addressed. For this aim, it provides a short overview of the current applications of IR systems in the context of genocide-related information, followed by a discussion of the ethical challenges of its human curation using a three-part framework inspired by Belmont criteria (i.e. respect for individuals, beneficence and justice/fairness). Finally, the chapter discusses to what degree these challenges are applicable to AI-driven IR systems and what can be the potential ways of bridging AI and memory ethics in this context.

## 2 AI-driven IR Systems and Genocide-Related Information

The digitisation of historical collections, together with the production of new digital-born materials dealing with information about genocides (e.g. audiovisual tributes to the Holocaust [25, 26]), prompted the growing use of IR systems within heritage institutions. Many of these systems partially reproduce or enhance the traditional curation practices used by archives or museums, but in some cases, IR systems substantially transform the scale and functionality of these practices.

For instance, Liew [14] examined how IR systems facilitate the exploration of collections, including the ones dealing with the Holocaust (e.g. by enabling keyword/phrase search and the use of wildcard operators). Schenkolewski-Kroll and Tractinsky [13] discussed the relationship between IR systems and authority lists in the context of Holocaust materials in the Israeli archives. Daelen [27] looked at the possibilities of using IR to enrich the inventory of collections related to the Holocaust and connect archives and users in the context of European Holocaust Research Infrastructure; similarly, Carter et al. [28] discussed the potential of AI-driven IR solutions for facilitating exploration of primary sources in the context of the Holocaust by proving new possibilities for user interaction with the Morgenthau Diaries.

In addition to IR systems enhancing traditional curation practices, there are also examples of more innovative applications of these systems within heritage institutions. One example is the use of three-dimensional visualisation of Holocaust survivors retrieving audio recordings of the survivors' earlier comments in response to the user input. Sometimes referred to as holograms [29] or social robots [30], these systems are used by several Holocaust memory initiatives (e.g. New Dimensions of Testimony or Forever Project) and enable new possibilities to retrieve testimonies through the simulation of human-to-human conversation.

Similar to other experimental proposals concerning the use of AI-driven IR systems in the context of genocide remembrance (e.g. the concept of personalised virtual reality-enhanced interaction with information about the Holocaust for the Babyn Yar memorial [31]), the use of social robots as a form of curation of genocide-related information has attracted not only praise but also criticism. For instance, Walden [29] noted that novel approaches for the use of IR (e.g. in the form of Holocaust survivors' holograms) do not necessarily meet the expectations about reactualisation of the past for the audience, whereas Alexander [32] noted that some of these novel approaches require not only historical knowledge but also media literacy, thus risking to make information less accessible for certain groups.

Not only heritage institutions but also commercial platforms are increasingly relying on IR systems for curating genocide-related information. The availability of digital content related to genocides coming both from the institutional (e.g.Holocaust museums [33]) as well as non-institutional entities such as online influencers [34] or artists [25] resulted in the growing presence of genocide-related information on the online platforms. These platforms range from social media sites, such as Instagram [34] or TikTok [35], to web search engines, such as Google [36], to commerce-related platforms (e.g. TripAdvisor [37]).

Under these circumstances, IR systems become a crucial element of genocide-related information curation, in particular, considering that commercial platforms often lack human curation expertise in this specific domain that differentiates them from heritage institutions. However, the implications of IR-driven curation currently remain unclear. Makhortykh et al. [36] examined how six web search engines curate visual information about the Holocaust and observed substantial differences in what aspects of the Holocaust are prioritised by the individual engines. Devon and Tobias-Hartmann [35] discussed the impact of the TikTok algorithm on the treatment of user-generated content, including content dealing with Holocaust denial, and found the tendency of the algorithm to suppress certain forms of resistance to antisemitic ideologies. Finally, Kansteiner [38] looked at how Holocaust institutions use IR systems associated with commercial social media sites (e.g. Facebook) and found the varying degrees of visibility of specific types of genocide-related content received.

Unsurprisingly, the use of AI-driven IR systems by commercial platforms for curating information about genocides also raised a number of concerns. In addition to the general critique of it undermining the gatekeeping functions of heritage institutions [39], studies suggest that IR systems used by platforms can promote factually incorrect or denialist content [36, 40]. Another concern relates to the possibility of commercial platforms' IR systems resulting in unequal treatment of information about different aspects of specific genocides (e.g. in terms of prioritising content coming from a few Holocaust sites while omitting the other ones [36]).

## 3 Memory Ethics and Human Curation of Genocide-Related Information

The major challenge of bridging AI and memory ethics in the context of genocide-related information curation deals with the multiple forms the curation might take. There are many approaches to human curation of such information, ranging from the one happening in heritage-focused environments, for instance, museums or archives, to wider public-focused environments, such as mass media.

The multiplicity of forms of curation prompts the importance of identifying which of them are particularly applicable to the discussion of AI-driven IR systems. While it can be debated, the argument can be made that human curation in archives is the closest in its nature. Both archives and IR systems determine what content is made visible to the public and what content remains hidden [23]: in the case of archives, these decisions are implemented by providing or not providing physical access to the collections, whereas in the case of IR systems, some outputs in response to user queries can be filtered out or down ranked.

Human curation of genocide-related information in the archival context has to deal with multiple ethical challenges [21, 24]. Some of these challenges are applicable to archival research in general, for instance, the potential damages to individual privacy [23]. However, other challenges are more specific to the case of genocide and include, for instance, the possibility of using archives to subjugate knowledge about the past atrocities [41] or impeding the processing of genocide-related trauma by encouraging specific types of testimonies and silencing others [42]. The need to address these challenges stimulates the discussion of how these ethical challenges can be addressed.

One of the common reference points in the discussion of ethics regarding human curation of archival information is Belmont criteria. Introduced at the end of the 1970s to provide guidelines for research involving human subjects, Belmont's criteria focus on three ethical principles: respect for individuals, beneficence, and justice (sometimes also referred to as fairness [24]). The recommendations of Belmont criteria generally suggest that "informed consent be sought, that benefits and risks be evaluated, and the selection, representation, and the burden of participation be fair and equitable" [43, p. 139].

A number of studies have critically interrogated to what degree Belmont criteria are applicable for archival research [22, 24]. Some studies argued that Belmont criteria are not applicable to archival research because it is fundamentally different from the other disciplines working with human subjects [23], whereas others (e.g. [24]) suggested that the criteria are focused primarily on preventing potential damage for the living subjects. However, in the case of genocide-centred research, many subjects are already dead that makes it hardly possible to obtain their consent for being involved in the research and the different set of risks/threats (e.g. potential damage to posthumous dignity of victims [17]) which have implications for the beneficence and justice criteria. Under these conditions, direct application of Belmont criteria to archival research dealing with genocides may undermine the ethical mandate of the genocide-focused scholarship [22].

Despite the above-mentioned drawbacks, it can be argued that Belmont criteria are still applicable for identifying the ethical challenges involved in the human curation of archival information about genocides. Specifically, this chapter proposes to apply a three-part framework using Belmont criteria—i.e. respect for individuals, beneficence and justice/fairness—to group together potential ethical challenges associated with genocide-related information curation. The rest of the section is devoted to the discussion of the individual challenges associated with each of the three criteria.

In the case of respect for individuals, it is possible to identify three major ethical challenges related to the human curation of genocide-related information: consent, double vision, and privacy. The first of these challenges—i.e. the need to acquire consent—is common for curation of information coming from human subjects in other contexts. However, in the case of genocide or other forms of mass violence, making sure that consent is acquired becomes a much harder task. In some cases, the difficulties can be due to substantial risks for witnesses or victims preventing them from voluntarily sharing information [44] or evidence being produced against the will of the victims [45].

Furthermore, the digitisation of genocide-related information raises additional questions such as, for instance, whether the consent given for the generation of analogue materials also automatically applies to their digitalisation and whether the difference between analogue and digital public access has implications for the consent to make information about the genocide publicly available [20]. These questions are particularly applicable for the historical instances of genocide (e.g. the Holocaust), where materials (e.g. testimonies) were produced in certain formats which have since then become outdated, so preserving them in the original format is both non-sustainable and ineffective from the point of view of communicating information about the genocide.

Another challenge of human curation relates to the problem of double vision, which relates to the transformation of genocide victims into objects (and not subjects) of research due to the distancing involved in the process of data collection and analysis [46]. Originally discussed in the context of processing analogue materials [22, 46], the problem of potential dehumanisation and depersonalisation of victims is amplified by the shift towards digital collections, enabling new possibilities for "anonymizing, numbering, and classifying" [24, p. 531] experiences of genocide victims as well as "converting humans into numbers" [47, p. 322].

One more aspect of respect for individuals concerns the matters of privacy. Archives, in general, can be damaging to the reputation of individuals whose information is disclosed without their consent [20, 23]. However, in the case of genocide, in particular its recent instances, privacy can be a matter of life and death, for instance, when either perpetrators or victims want to take revenge in their hands. At the same time, the profound anonymisation of genocide-related records has been criticised for its potential for erasing the voice of victims [22], which in some way is similar to the purpose of the genocidal actions aiming to erase any traces of victims.

For the beneficence of the human curation of genocide-related information, it is possible to identify two major challenges: the problem of representation and the possibility of distortion/manipulation. The former challenge relates to the argument that because genocides are instances of unprecedented violence, any attempt of their representation (e.g. via certain modes of storytelling or documentation [30, 48]) is inadequate. Hence, the absence of representation might be a "more accurate or truthful or morally responsive" [49, p. 71] way of dealing with genocide-related information.

The second challenge concerns the possibility of information about genocide being distorted or manipulated. The forms of distortion of historical information can vary broadly; some examples include de-contextualisation of historical phenomena [20], denial or justification of the past crimes [50] or the use of references to past suffering or injustice for stigmatising specific social groups in the present [19]. In the case of genocide-related information, such forms of distortion are particularly concerning both due to the ethical obligations of protecting the memory of victims and the strong affective potential of information about past injustices, which can be used to incite violence in the present [26].

Finally, the justice/fairness of human curation concerns two interrelated aspects: the politicisation of curation and the unequal treatment of specific types of genocide-related information. One of them is the politicisation of archives, which has implications for what information about the past atrocities is available and how it is communicated to the public [24, 47]. The transmission of the matters of curation of genocide-related information to the realm of politics might not only downplay the importance of ethical obligations associated with it but also facilitate instrumentalisation of genocide memory for immediate political gains. Such instrumentalisation can lead to genocide-related information being used to manipulate public opinion, for instance, to justify violence in the present, as it happened in the case of the Russian aggression against Ukraine.

The second justice-related challenge deals with the unequal treatment of certain types of genocide-related information. In some cases, it is attributed to politicisation of archives, which can lead to the release of information (e.g. in the form of archival documents) supporting certain political agendas [47] or silencing of information which can be viewed as damaging for a ruling regime [41]. In other cases, the unequal treatment can be related to the belief that some types of information can be less reliable (e.g. due to the assumption that genocide victims can not provide a neutral view on the genocide [21]) or the imbalance between the availability of different types of information (e.g. because of certain groups of individuals being more likely to survive and, thus, leave testimonies [47]).

**4 Bridging Memory Ethics and AI-driven IR System Design**

After identifying the common ethical challenges of human curation of genocide-related information, it is important to examine to what degree they are applicable for IR system-based curation and how IR system design can be bridged with memory ethics to address concerns associated with these challenges. For this purpose, this section will use the same framework of three groups of challenges related to respect for individuals (consent, double vision, and privacy), beneficence (representation and distortion/manipulation), and fairness/justice (politicisation of curation and unequal treatment).

In the case of consent, the difficult part of bridging ethics and IR design relates to consent granting usually being part of the initial stage of data generation (e.g. the recording of a testimony or registering of an account to upload digital materials). One exception here relates to IR systems used in the context of web search, where indexing of digital-born materials is an ongoing process. However, in most cases, the IR systems process data for which consent has already been granted (e.g. in the case of collections stored in the heritage institution or materials generated through the online platform). While it can be possible to integrate consent checks or regular requests for consent re-granting, this specific aspect can arguably be more relevant for the overall model of institution/platform functionality and not necessarily for the IR design.

In terms of double vision, AI-driven IR systems are sometimes argued [30] to be capable of encouraging trust and empathy, which can counter the potential dehumanisation of victims associated with this challenge. Such a problem can be particularly pressing with the passing of the living witnesses of the genocides, which amplifies the risk of them being increasingly treated as objects and not the subjects of research. One particular example of the use of IR for countering this issue is the shift towards more human-to-human communication-like forms of IR, for instance, the use of conversational agents as a form of curation of genocide-related information.

The potential of IR systems for dealing with privacy-related challenges shares certain similarities with the case of content. In some cases, the matters of privacy are dealt with during the initial state of data generation (e.g. in the case of thorough anonymisation of genocide-related evidence). However, in other cases, IR systems can have a rather ambiguous impact on the privacy of individuals the information about whom they are curating. Advancements in several fields of AI (e.g. computer vision and natural language processing; [63]) enable novel possibilities for recognising the presence of individuals or mentions of specific entities in the data; however, the same advancements can be used for protecting individual privacy (e.g. by masking private information present in the documents).

Under these circumstances, the inclusion of particular functionalities in the IR system design can either expose or protect private information. The choice of functionalities can be informed by examining the selection of materials which the system is expected to work with and its intended uses.

Filtering out privacy-sensitive content can be another alternative to making modifications to the original data. Often associated with the right to be forgotten [58], also known as the right to erasure, this approach can be particularly applicable in the case of IR systems dealing with genocide-related information in the context of web search, where the modification of indexed data is not necessarily possible. Applied for protecting individual privacy, this mechanism might be less applicable for genocide-related information, where its evocation for specific cases (e.g. the application of the right to be forgotten for perpetrators of genocide) might contradict the public interest [59]. However, in other cases (e.g. protecting the victims), it might be essential for tackling privacy-related challenges, thus stressing the importance of functionalities that can facilitate requests for the activation of the right to be forgotten in the IR system design.

From the point of view of beneficence-related challenges, IR can enable new possibilities for addressing the problem of representation. The new formats of curating genocide-related information (e.g. via social bots [30]), as well as more personalised approaches (e.g. the ones taking into consideration the level of knowledge identified on the basis of earlier history of interactions with the IR system), can move the genocide-focused storytelling beyond the traditional modes of representation. While the adequacy of these novel IR approaches for dealing with genocide-related information can be debated, it is important to investigate their potential.

Similar to addressing the problem of representation, AI-driven IR systems can facilitate countering distortion of genocide-related information. The possible approaches for doing it vary from automated detection of distorted information (e.g. the denialist claims) and their subsequent filtering/de-prioritisation (e.g. in the case of Holocaust denial content being countered by commercial platforms) to the provision of contextual information to the system outputs dealing with the genocide. The growing body of research on integrating mechanisms of detecting and countering misinformation in AI-driven IR systems [60, 61] demonstrates possibilities provided by these systems for preventing the distortion of historical facts.

At the same time, it is important to acknowledge the dangers posed by IR systems to the beneficence of curation of genocide-related information, in particular, in the context of the increasing complexity of IR systems [30]. Such complexity makes it more difficult to identify potential instances of system manipulation, in particular, for the users having a limited understanding of the logic behind the system functionality. Together with the limited knowledge about the overall composition of the pool of outputs (e.g. in the case of web search IR systems dealing potentially with billions of genocide-related outputs), it stresses the importance of integrating transparency in the IR system design.

In the case of concerns about archives' politicisation, the impact of AI-driven IR systems can be ambiguous. Depending on how aware of the functionality of IR systems actors involved in politicisation of genocide-related information are and to what degree these actors are capable of influencing the system, IR systems can either facilitate politicisation or counter it. Contextual factors are particularly important for instance, under the condition of intense politicisation of genocide-related information within a particular country, the transparent functionality of IR systems used by the local heritage institutions may actually facilitate the appropriation of the systems for controlling information curation. By contrast, non-transparent IR mechanisms used by a transnational company that is less dependent on the whims of the local memory regime may actually counter politicisation by offering a less politicised selection of information.

The question of the intended use of the IR systems is also of particular importance for identifying their ability to deal with unequal treatment of genocide-related information. Similar to IR systems dealing with news [62], genocide-focused IR systems can serve different normative functions. More deliberative models of IR systems can be optimised for enabling equal representation of genocide-related information (e.g. in terms of visibility of specific aspects of the genocide or particular sites [36]) via either personalised or non-personalised curation, whereas more liberal models might omit the matters of equality, instead giving visibility to a few prominent aspects which the system expects the user to be particularly interested in. The preference for a particular model determines the logic behind the design of a particular IR system; however, determining such a preference might be a rather non-trivial task (e.g. what stakeholder groups shall be able to decide on it?), which is also true for realising more complex models of information curation (e.g. what characteristics to take into consideration when deciding on the equality/lack of equality in representation of specific aspects of a genocide?).

## 5 Discussion

The chapter scrutinised the ways for bridging AI and memory ethics in the context of IR systems dealing with genocide-related information. Using a Belmont criteria-inspired typology of ethical challenges associated with human curation of information about genocides, it discussed to what degree IR systems can address curation issues related to respect for individuals, beneficence and justice/fairness. The results of this discussion highlight several important points concerning the potential of IR systems for curating information about genocides, both historical (e.g. the Holocaust) and recent ones (e.g. Rohingya genocide).

The first point suggests that AI-driven IR systems are, unfortunately, not a silver bullet capable of easily solving ethical challenges associated with the curation of genocide-related information. Even while they can address some of the issues related to human curation (e.g. by enabling new possibilities for addressing some of respect- or fairness-related challenges), they can also worsen other issues (e.g. the beneficence-related challenges), in particular, considering the high complexity and frequent lack of transparency of IR systems. Under these circumstances, it becomes of paramount importance to take into consideration the complex relationship between IR systems and memory ethics when designing the former to minimise the possibility of IR having detrimental effects on the lives of individuals affected by genocides and on genocide remembrance.

Second, similar to other domains (e.g. journalism [51]), there is a tradeoff between the realisation of AI potential for enhancing the performance of IR systems (e.g. in terms of addressing ethical challenges, in particular, the ones related to beneficence and justice/fairness) and transparency. While the increased complexity of IR systems enables new possibilities for making the treatment of different groups of genocide victims more fair (e.g. in terms of making their suffering equally visible via AI curation) and dealing with the problem of representation (e.g. in terms of filtering out and removing information distorting historical facts), it also makes the functionality of these systems less transparent, thus limiting the user control over the system [52].

Third, the growing presence of genocide-related information on commercial (and not only heritage-oriented) platforms poses additional difficulties for its curation through IR systems. Because of the generalist focus of commercial platforms (e.g. Google), it is difficult (albeit not impossible, as shown by the case of COVID-related information moderation [53]) to enable distinct treatment of specific types of information. Under these circumstances, IR systems used by commercial platforms often treat information about sensitive and traumatic subjects (e.g. genocides) in the same way and follow the same logic (e.g. to maximise user engagement as in the case of some social media sites) as other subjects such as entertainment topics. The possibility of such non-differentiated treatment can result in a number of ethics-related issues (in particular, related to the respect for individuals and beneficence of curation) and prompts the importance of the dialogue between the commercial platforms and heritage practitioners as well as other genocide-related actors (e.g. survivors or their families) in order to find a way for addressing these issues.

Finally, it is important to note several limitations of the conducted study. The primary limitation is the reliance on a conceptual approach to discuss the relationship between IR systems and memory ethics. Specifically, the chapter relies on the existing academic scholarship for synthesising the main challenges of human curation of genocide-related information and discussing the possible ways of addressing them through using AI-driven IR systems. Future research can benefit from a more empirically-driven approach (e.g. based on interviews) to solicit opinions of heritage practitioners on the ethics-related issues involved in curation of information about different genocides as well as how these can be affected by IR systems.

A related challenge concerns the focus on the existing research on memory ethics and information curation regarding one particular instance of genocide, namely the Holocaust. While such a focus is not surprising considering the particular importance of the Holocaust, in particular, for the Global North [54], it is crucial to acknowledge that information about other genocides might pose different challenges, in particular considering the uniqueness of each genocide [55] as well as the increasing criticism of West-oriented standardisation of genocide commemoration [56, 57]. Under these circumstances, it is important not only to extend the discussion of the role of IR systems to other instances of genocide, including the ones occurring in the Global South and Global East, but take into consideration that requirements for AI-driven IR systems in these cases may be different.